# Interlayer coupling effect in twisted stacked few layer black phosphorus revealed by abnormal blue shifts in Raman spectra

Tao Fang[1, 4], Teren Liu[2], Zenan Jiang[3], Rui Yang[2], Peyman Servati[3] and Guangrui (Maggie) Xia[2,*]

[1]Department of Physics and Astronomy, University of British Columbia, Vancouver, BC, Canada, V6T 1Z4

[2]Department of Materials Engineering, University of British Columbia, Vancouver, BC, Canada, V6T 1Z4

Tel: +1-604-8220478, Fax: +1-604-8223916, E-mail: guangrui.xia@ubc.ca

[3]Department of Electrical and Computer Engineering, University of British Columbia, Vancouver, BC, Canada, V6T 1Z4

[4]Stewart Blusson Quantum Matter Institute, University of British Columbia, Vancouver, British Columbia V6T 1Z4, Canada

## Abstract

In this work, twisted stacked few layer black phosphorus heterostructures were successfully fabricated. Abnormal blue shifts in their $A_g^1$ and $A_g^2$ Raman peaks and unique optical reflections were observed in these samples. The phonon behavior difference can be explained by our density functional theory calculations, which suggest that interlayer coupling has a significant effect in twisted bilayer black phosphorus. According to the calculations, the interlayer interactions are not simply van der Waals interactions. Additional interactions, such as weak valence bonding between the top and bottom flakes, are considered to be the cause of the blue shifts in their Raman spectra.

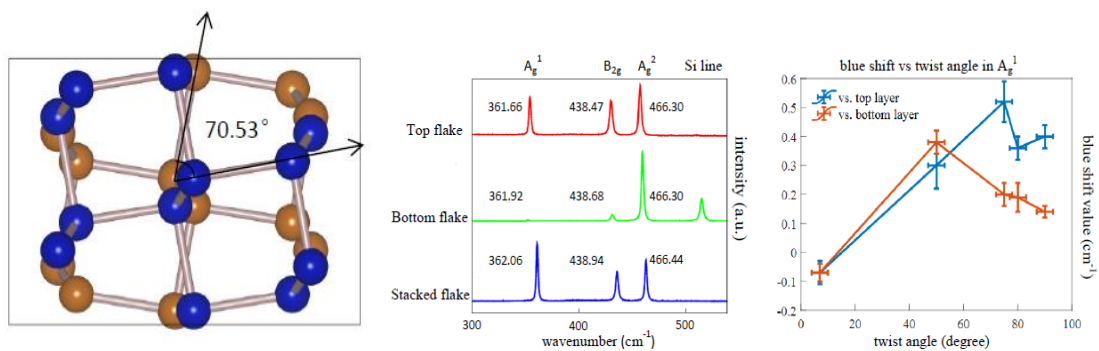

**Introductions**

Two dimensional (2D) materials have attracted much attention in recent years, due to their unique properties and wide applications. [1] 2D black phosphorus (BP), first isolated in 2013, is a good candidate for flexible electronic and photonic devices. [2] Like other 2D materials, BP is composed of 2D atomic layers and the interlayer interactions are commonly considered to be weak van der Waals type.

When two identical 2D structures are superposed with a twist angle θ, this new structure, called Moiré pattern, may generate interference, change electron properties and activate new phonon modes. These changes can be examined by optical observation and Raman spectra. New Raman peaks or shifts of Raman peaks are expected. [3, 4] Stacked 2D structures were first studied in graphite and graphene [5], and later in transition metal dichalcogenides including molybdenum disulfide ($MoS_2$) and molybdenum diselenide ($MoSe_2$) [3-7]. Previous studies have shown that for twisted bilayer graphene, new Raman peaks appear depending on the twist angle and the excitation laser [3]. Also, optical reflection changes by changing the twist angle, which leads to 'microscopic rainbows' in twisted bilayer graphene [6, 7]. For $MoSe_2$, a shift in interlayer breathing modes and shearing modes were observed [4]. As a result, previous study indicated that significant interlayer interactions in 2D materials exist, which strongly depend on the interlayer twist angle [8, 9].

For 2D BP, previous theoretical studies predicted that twisted bilayer BP has unique band structure and optical response, different from untwisted bilayer BP because of interlayer interactions. For example, it has unique absorption spectra, which makes it useful as a filter material [10]. Also, theoretical calculations predicted that its electronic structure and optical transitions were tunable by external gating [8]. Other band structure simulations were consistent with this conclusion, and pointed out that the twist angle is a new degree of freedom in this superlattice [11]. However, these effects have not been observed experimentally, because 2D BP is delicate [12], difficult to transfer and not stable in air. Moreover, BP made from chemical vapor deposition (CVD) is not sufficiently crystalline [13].

Wet transfer methods and dry transfer methods are commonly used to fabricate 2D heterostructures [14, 15]. However, these methods are not suitable for making twisted few layer BP without careful considerations or modifications. An alkali solution is essential in wet transfer, which oxidizes BP. Polydimethylsiloxane (PDMS), which is used commonly in a dry transfer process, is too sticky such that BP flakes on a PDMS film tend to stick on it, rather than stacking to another BP flake. In this work, we figured out a modified dry transfer method, and was able to fabricate twisted stacked few layer BP successfully.

**Experiments**

After mechanical exfoliation, we used a modified dry transfer method to fabricate twisted stacked BP. A chloroform solution with 6% polycarbonate (PC) was dropped on a glass slide to make a PC film. After that, BP flakes were transferred to this film by dry transfer. Optical microscopy was used to select proper BP flakes of sufficient sizes and uniformity on the PC film to be further transferred. Then, we positioned the PC film with BP flakes on a BP flake bearing silicon substrate by a micro-feeding unit. In this step, the PC film/glass slide was in contact with the Si substrate face-to-face. The Si substrate was mounted on a heating stage. It was then heated to 190 °C for five minutes to make the PC film crack and peel off from the PC/glass slide. Finally, the PC film was removed by putting the sample in the chloroform solution for one minute. This transfer method may be called as "PC film based transfer". This method provides a convenient way to transfer BP on other 2D materials, and BP/graphene heterostructure were successfully fabricated by this method.

So far, theoretical calculations and simulations mainly focused on twisted bilayer BP, which is composed of two single layer BP flakes. However, it is difficult to fabricate two stacked monolayer BP experimentally. BP is well known to oxidize easily in air [16]. It is hard to obtain monolayer BP crystals by pure mechanical exfoliation [2, 17], and BP made from liquid phase exfoliation has a comparatively small size (less than 5μm) even with a size selection method [18, 19]. Current thinning methods for BP, such as thermal sublimation, plasma thinning and local anodic oxidation with water rinsing, are still not sufficient to ensure a stable outcome [20-22]. Therefore, in our experiment, we could only fabricate twisted few layer BP, or twisted stacked bulk BP, which is different from twisted bilayer graphene and molybdenum diselenide, which are composed of two single atomic layers.

**Density functional theory (DFT) calculations**

DFT calculations of twisted and untwisted bilayer BP were conducted to provide theoretical insights on the impact of the interlayer coupling. The detailed construction of twisted bilayer BP superlattice was conducted by D. Pan et. al [23]. The new lattice base vectors could be constructed by transformation as the following:

$$\begin{pmatrix} b_1 \\ b_2 \end{pmatrix} = \frac{a}{2} \begin{pmatrix} -sin\varphi & cos\varphi \\ sin\varphi & cos\varphi \end{pmatrix} \begin{pmatrix} i \\ j \end{pmatrix} \quad . \tag{1}$$

Here, i, j are basis vectors for untwisted BP. φ is the angle between two basis vectors of untwisted BP lattice, which is 35.27 °. $b_1$, $b_2$ are the new basis vectors for twisted bilayer BP superlattice.

For pure twisted stacking without slipping, the Moiré twist angle is given by [24, 25]

$$\cos\theta = \frac{(q^2+p^2)cos2\varphi+q^2-p^2}{q^2+p^2+(q^2-p^2)cos2\varphi} \quad p,q \in N_+ \quad (2)$$

Simple unit cells and thus superlattices only exist with discrete p and q values. The simplest superlattice corresponds to p = q = 1. In this case, the twist angle is 70.53° between the layers. On each atomic layer, the area occupied by a unit cell of twist bilayer BP is three times that of untwisted bilayer BP. Since there are two atomic layers, altogether there are 4×3×2=24 phosphorus atoms in one unit cell (Fig. 1b). Therefore, there are be 24×3=72 phonon modes in this superlattice.

The DFT calculations were based on the simplest unit cell and superlattice discussed above. Even for this simplest case, the calculations of phonon modes were quite time consuming (see *Methods* for details). For comparison, we calculated Raman spectra and phonon modes for untwisted bilayer BP. Due to time and computing resource constraints, the phonon modes of more complicated superlattices were not calculated.

**Results and Discussions**

Five twisted stacked BP structures were fabricated with different twist angles, which are 7, 50, 75, 80 and 90 degrees respectively. The optical image of a twisted few layer BP sample is shown in Fig. 2. The top and bottom flakes are 20 and 8 nm thick respectively, and the twisted angle is 90±3 degrees measured by angle resolved polarized Raman spectroscopy. The color of the overlapping area is obviously different from the two non-overlapping areas. According to the color-thickness correlation for untwisted BP [26], the color should be light green for untwisted BP with a thickness 28 nm, but the actual color of the overlapping area with a twist angle of 90° is dark green with a twist angle of 90 degree. The color difference was observed in three more twisted BP samples with twist angles of 50°, 75° and 80°. The color difference between a sample with a 7° twist angle and its untwisted counterpart is insignificant. The color changes indicate that the optical reflections of twisted BP flakes are different for the untwisted BP, which reflect the changes in electron properties and bond structure. This presumption agrees with previous simulations in bond structure [23].

Experimentally, high frequency Raman spectra of BP have three significant Raman peaks, which are $A_g^1$, $B_{2g}$ and $A_g^2$ modes. The corresponding vibrations are illustrated in Fig. 3 (a). From the Raman spectra in Fig. 3 (b), $A_g^1$ and $A_g^2$ Raman peaks of the overlapping area show blue shifts of 0.4 and 0.2 cm$^{-1}$ respectively compared with those of the non-overlapping areas. $B_{2g}$ peak is blue-shifted in this sample, but it may red-shift in other samples, depending on the twist angle. Previous studies have shown that $A_g^1$, $B_{2g}$ and $A_g^2$ modes should be red-shifted in untwisted BP on $SiO_2$/Si substrates when thickness increases [17, 27], due to a laser heating and the sample heat absorption and conduction difference. In our experiment, the overlapping area is thicker than non-overlapping area, but the Raman peaks observe a blue shift, rather than a red shift. Because of that, we call this phenomenon "abnormal blue shift". Other influence factors, such as laser heating effect and equipment errors (+/-0.05 cm$^{-1}$) were excluded by careful examination. Discussions related to these factors are in the supplement information.

The twist angle dependence of the blue shift for the three Raman peaks is shown in Fig. 3 (c) to (e). The Raman peaks of the sample with 7° twist angle show red shifts with the increased thickness, which suggests that the phonon modes of a stacked BP with a small twist angle is close to an untwisted stacked BP. Excluding this sample, for the out-of-plane vibration mode $A_g^1$, the peak shifts are all blue shifts of 0.2 to 0.5 cm$^{-1}$; for in-plane vibration mode $A_g^2$, a similar trend was observed with smaller blue shifts and for in-plane vibration mode $B_{2g}$, the peak shifts measured are more complicated showing one peak being red-shifted and others blue-shifted.

The DFT calculation results are summarized in Table 1. The unit cell of the twisted bilayer BP is in space group P222 (point group $D_2$), while the unit cell of the untwisted bilayer BP is in space group P*bcm* (point group $D_{2h}$). Due to the lower symmetry, or "symmetry breaking", the 72 calculated phonon modes of the twisted bilayer BP with 70.53° angle (24 atoms per unit cell) do not have special parity, and the symmetry notations reduce to A and B instead of $A_g^1$, $B_{2g}$ and $A_g^2$. Out of the 72 modes calculated, to identify the corresponding Raman peaks of twisted bilayer BP, we select two high intensity Raman active modes with same symmetry notation A, whose wavenumbers are close to the wavenumbers we measured in the experiment. To make the discussion easier to understand, in the following, the measured and calculated Raman peaks of twisted stacked BP are still called $A_g^1$, $B_{2g}$ and $A_g^2$ peaks. For example, the measured $A_g^1$ and $A_g^2$ peaks were at 362.06 and 466.44 cm$^{-1}$ for twisted BP as in Fig. 3(b) and the corresponding peaks selected from the DFT results of twisted bilayer BP are at 356.56 and 456.88 cm$^{-1}$. The relative differences of the wavenumbers are less than 3%. The detailed list, for all the 24 phonon modes in untwisted bilayer BP and 72 phonon modes in twisted bilayer BP, is in the supplement information.

Next, we compared the calculated results of an untwisted bilayer BP and a twisted bilayer BP with a 70.53° twist angle. DFT calculation results of $A_g^1$ and $A_g^2$ peaks of twisted bilayer BP are 4.70 cm$^{-1}$ and 1.31 cm$^{-1}$ higher than those of untwisted bilayer BP, which are consistent with the abnormal blue shift observed in Raman spectra. Another interesting fact is that the shift values are larger in calculation than those in experiment. This may be explained by the BP layer thickness differences between the experiments (few-layer) and the calculations (bilayer). Thicker flakes may reduce the phonon modulation effect and decrease the blue shift value.

Charge densities in twisted and untwisted bilayer BP were also calculated by DFT program. The results are compared in Fig. 4. The 3D charge distribution of the twisted bilayer BP in Fig. 4 (b) has more overlap between the two layers, while that of the untwisted one in Fig. 4 (a) is more localized. The cutting planes (020) in Fig. 4(c) and (d) are the middle-distance planes between the two layers. Significant distribution differences exist between two figures. A weak valence bond, as indicated in the graph, may exist in twisted bilayer BP. Another cutting plane (100) of twisted bilayer BP in Fig. 4(e), which is normal to the 2D atomic planes, gives another perspective of valence bond between phosphorus atoms in different layers.

Based on the experiment and calculation results, this abnormal blue shift can be explained by a strong interlayer coupling effect and thus a strong phonon mode change near Γ point in this 2D heterostructure. This is consistent with the significant interlayer interaction reported before [9]. A phosphorus atom has five valence electrons, three of which form covalent intralayer bonds and the rest two form electron lone pair. This electron lone pair may extend to the interlayer vacuum region and affect interlayer van der Waals interactions. When the twist angle is not equal to zero, there are different Coulomb interactions between layers because of electron distribution change. This change will cast a noticeable effect on phonon modes and cause abnormal blue shift in Raman spectra. $A_g^1$ mode has a vibrating direction perpendicular to the 2D BP plane, so it has a larger shift compared to other modes. This phenomenon also indicates that BP interlayer interactions are not simply van der Waals interactions. Weak valence bond, as indicated in Fig. 4, may exist between atoms in different layers of twisted bilayer BP. Other phenomenon reported before, such as surface reactivity, proves additional interactions may exist as well [28, 29].

More insights may arise from this work. Previous simulation predicted that the Dirac cone and topological phase transition may be resulted from external electric field in BP. Strong interlayer interactions are an indispensable part for this phenomenon [30]. Low frequency Raman measurements may be useful to find new peaks and more shifts in twisted stacked BP. This is a topic worthy more exploration.

**Conclusion**

In conclusion, twisted stacked few layer BP heterostructures were successfully fabricated. Unique optical reflections were observed in these samples, which also have abnormal blue shifts in their Raman spectra. The phonon behavior difference can be explained by our density functional theory calculations, which suggest that interlayer coupling has a significant effect in twisted bilayer BP. According to the calculations, the interlayer interactions are not simply van der Waals interactions. Additional interactions, such as weak valence bonding, are considered to be the cause of the different Raman spectra.

**Methods**

*Exfoliation*: Bulk BP was purchased from Smart Elements. Few-layer BP were mechanically exfoliated onto a Si substrate with 300 nm of $SiO_2$ layer. Before exfoliation, the silicon substrate was cleaned by acetone and IPA subsequently. Optical observation was used to preliminarily determine the position and thickness of BP. The detailed examination of thickness was carried by AFM measurement.

*Optical characterization*: Optical measurements were carried out by a Nikon TE2000 optical microscope, which was linked to an imaging spectrometer and a computer. A quartz-tungsten-halogen lamp was used to illuminate the BP sample. The magnification was 20 times. The color level in the computer image was set to be the same as vision.

*Raman spectroscopy*: Raman spectra were collected by a Horiba Scientific LabRam HR800 Raman system. The excitation laser is 441.6 nm in this experiment (2.81 eV, He-Cd laser). Between the detector and the filter, there was a polarization analyzer for angle-resolved polarized Raman spectroscopy (ARPRS). The spectral range measured was from 300 to 600 $cm^{-1}$.

*Twist angle characterization*: ARPRS was used to measure the twist angle. The intensities ratio of $A_g^1$ and $A_g^2$ are strongly dependent on the angle between the polarization direction of the incident laser and the crystal orientation of the sample [27, 31, 32]. In the ARPRS work, crystal orientations of non-overlapping and overlapping areas were measured simultaneously. BP samples were rotated by 15° each step with a total twenty four steps to cover 360°. We keep the position of BP sample the same in the whole process to avoid ambiguity and uncertainty.

*Atom Force Microscopy (AFM)*: AFM measurements were performed in a nano IR system. The mode was tapping mode with a resonance frequency 285 kHz. The scanning rate is 1 Hz.

*DFT calculation:* The construction of the Moiré pattern superlattice, as shown above, was based on the Coincidence Site Lattice (CSL) theory. Raman spectra of the twisted and untwisted bilayer BP were calculated within the framework of the DFT by Quantum Espresso software package [33]. Raman and infrared intensities were calculated by the phonon modes and second order derivatives of the electron density matrix in unit cell. PBE (John P. Perdew, Kieron Burke and Matthias Ernzerhof [34]) functional with norm conserving potentials was used, and the energy cutoff was set to 80 Ry. To increase the accuracy of the phonon calculations, the self-consistency thresholds were set to $10^{-14}$. The calculation was performed by 4 computing nodes with 32 cores and 128 GB memory on each node, and took about 20864 CPU hours.

**Author contribution**

Tao Fang and Guangrui Xia initiated the project. Tao Fang designed the experiments. Tao Fang and Rui Yang made BP samples. Tao Fang and Teren Liu carried Raman and AFM measurements. Teren Liu and Tao Fang did DFT calculations. Tao Fang, Teren Liu and Guangrui Xia led the writing and the revisions of the paper. The project was supervised by Guangrui Xia.

**Acknowledgement**

We would like to acknowledge these people for their help in this project: Prof. Mona Berciu (Department of Physics and Astronomy, University of British Columbia) for project initiation and constructive suggestions, Prof. Joshua Folk (Department of Physics and Astronomy, University of British Columbia) for providing experimental apparatus, Qian Song (graduate student at Department of Physics, Massachusetts Institute of Technology), Jialun Liu (Graduate student at Department of Electrical Engineering and Computer Science, University of Illinois at Urbana–Champaign) , Ursula Wurstbauer (Walter Schottky Institute) and Poya Yasae (graduate student at Department of Mechanical and Industrial Engineering, University of Illinois at Chicago) for useful discussions. Also, we would like to acknowledge Westgrid and Compute Canada, who supported our DFT calculations.

**Competing financial interests**

Tao Fang would like to acknowledge Stewart Blusson Quantum Material Institute (SBQMI) to provide funding for this project.

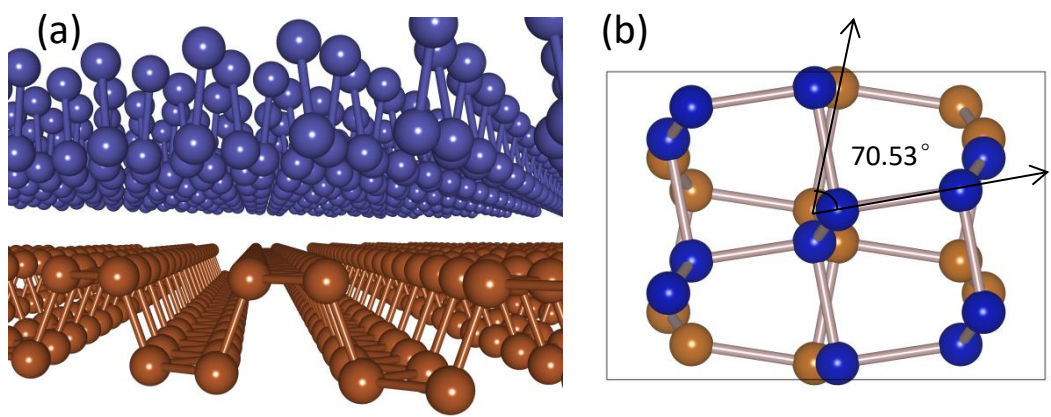

**Figure 1.** (a) Crystal structure of 70.53° twisted bilayer BP. (b) Unit cell of the 70.53° twisted bilayer BP. The projection direction is along b-axis (perpendicular to 2D BP plane). This is the simplest unit cell among those of twisted bilayer BP superlattices.

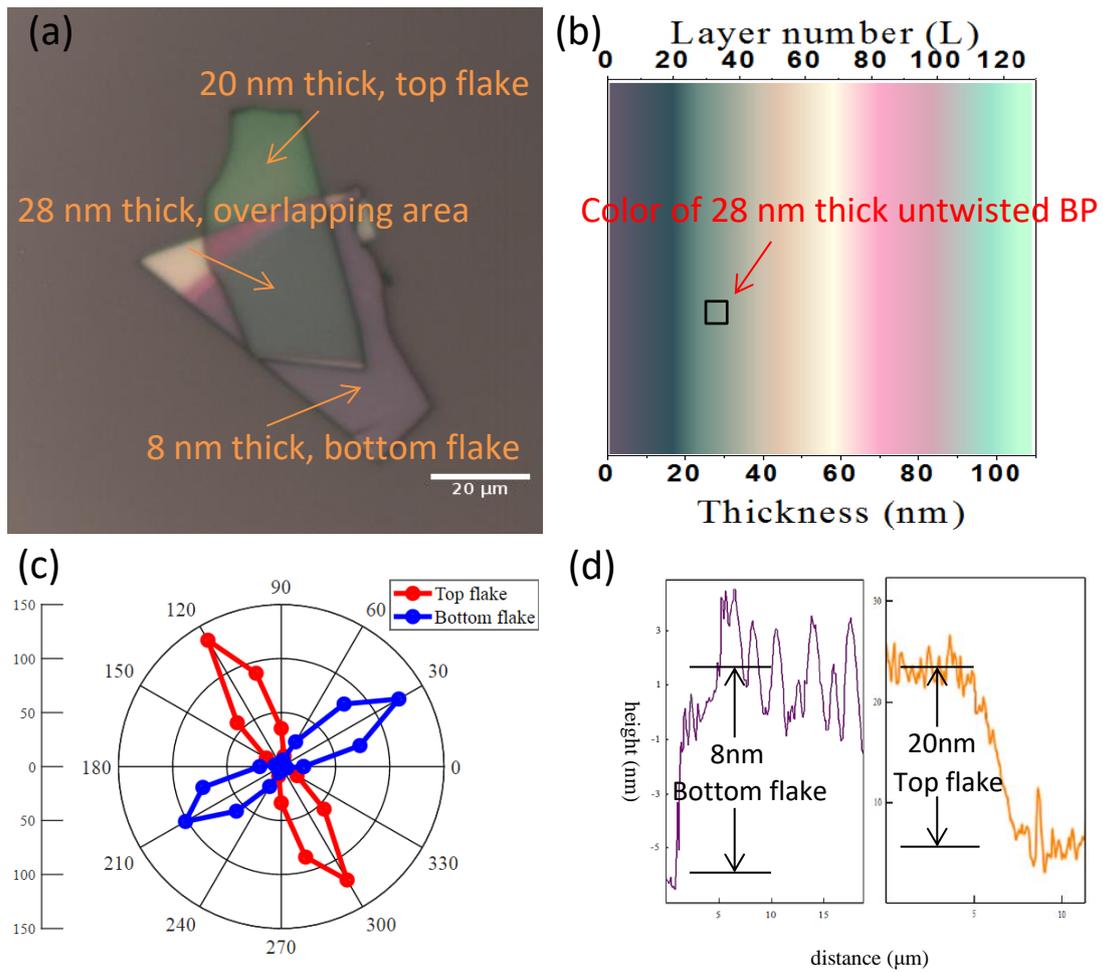

**Figure 2.** (a) Optical image of twisted few layer BP. (b) Color-thickness map of untwisted BP, and the thickness were obtained from AFM measurements. (c) Twist angle characterization by angle-resolved polarized Raman spectroscopy (ARPRS). Crystal orientations of top flake and bottom flake were measured simultaneously. The twist angle for this sample is 90±3 degrees. (d) AFM data of twisted few layer BP. The thicknesses of top and bottom flake are 20 nm (~20 layers) and 8 nm (~8 layers) respectively. The total thickness is 28 nm.

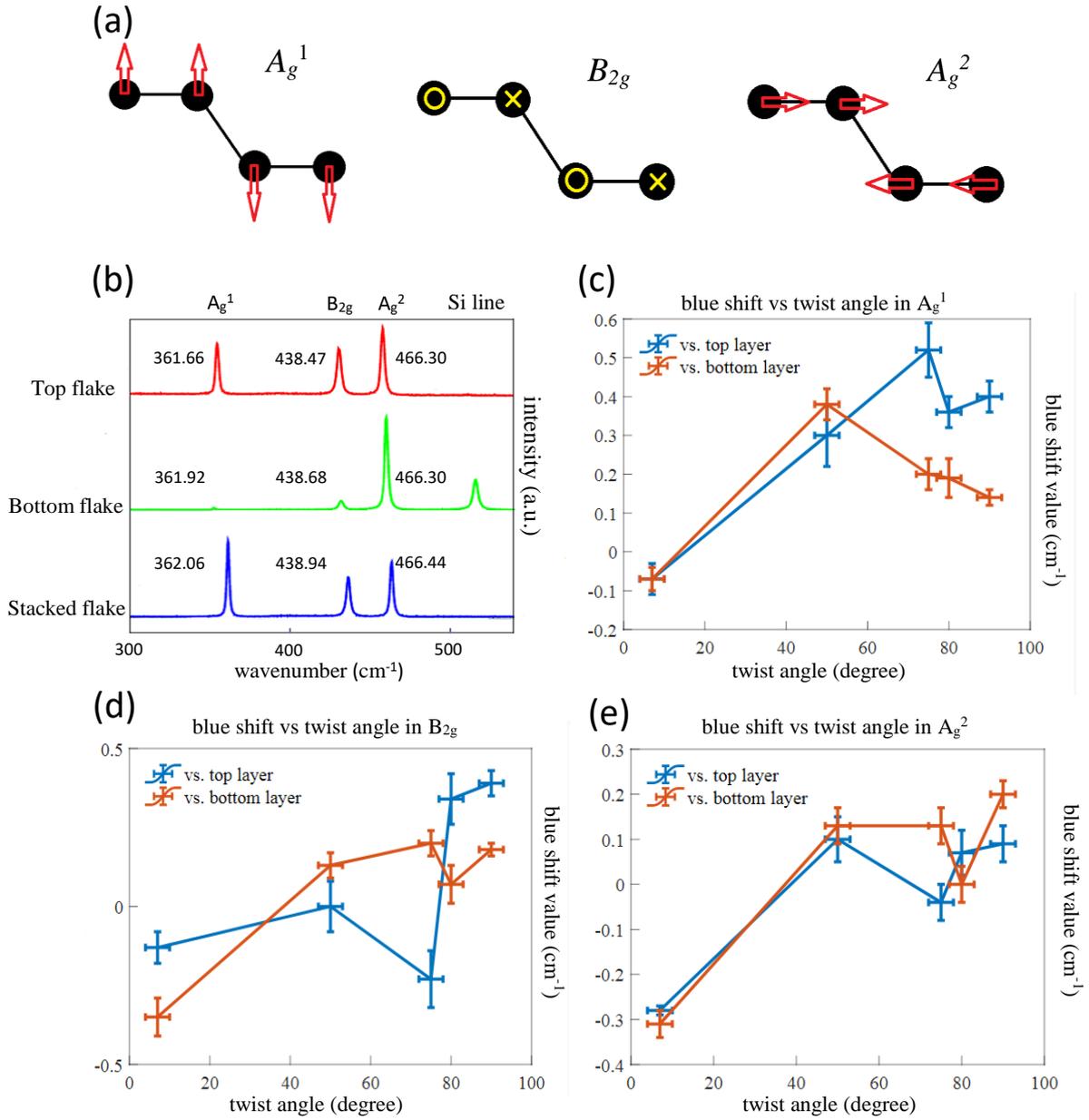

**Figure 3.** (a) Illustration of three high frequency Raman active phonon modes in BP. $A_g^1$ mode has a vibration direction perpendicular to 2D BP plane, so interlayer coupling exerts more influence on it. (b) The Raman spectra of top flake, bottom flake and stacked flakes in sample shown in Fig. 2 (a). The wavenumber of stacked flakes is larger than both top flake and bottom flake, so the Raman spectra of overlapping area is not the superposition of that in top flake and bottom flake. We call this phenomenon 'abnormal blue shift'. (c) - (e) Plot of blue shift value versus twist angle in $A_g^1$, $B_{2g}$ and $A_g^2$. Values of top layer and bottom layer are plotted separately. A positive value means a blue shift while a negative value means a red shift.

| Mode | Type of BP | Symmetry | Wavenumber (cm$^{-1}$) | Raman Intensity | Blue shift value (cm$^{-1}$) |
|---|---|---|---|---|---|
| $A_g^1$ | Untwisted bilayer | $A_g^1$ | 351.86 | 18768.502 | 4.70 |
| | Twisted bilayer | A | 356.56 | 26606.2044 | |
| $A_g^2$ | Untwisted bilayer | $A_g^2$ | 455.57 | 28592.0377 | 1.31 |
| | Twisted bilayer | A | 456.88 | 1655.471 | |

**Table 1** DFT calculation results of untwisted bilayer BP and twisted bilayer BP. These two modes are selected from the 24 phonon modes in untwisted bilayer BP and the 72 phonon modes in twisted bilayer BP with same kind of symmetry.

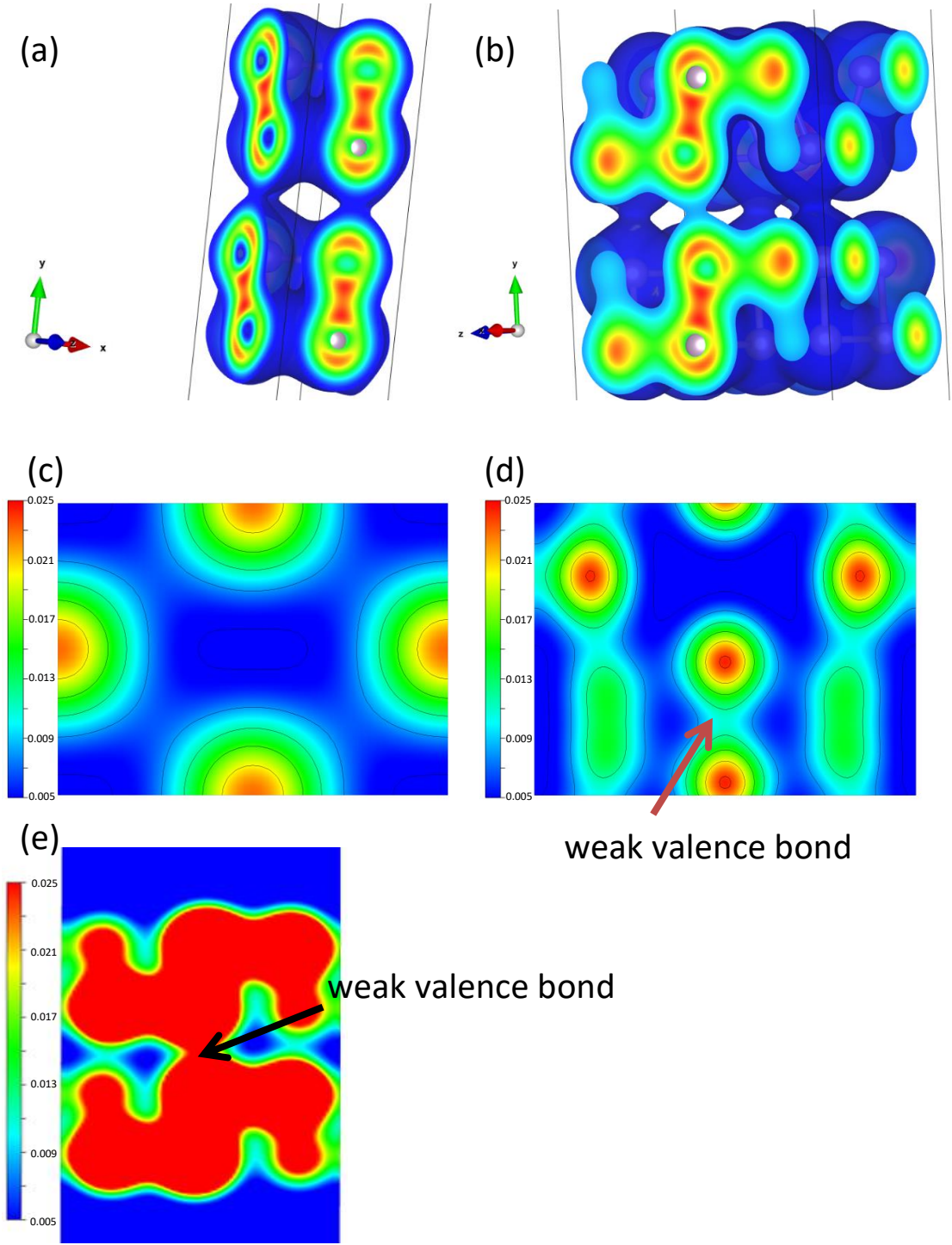

**Figure 4.** (a) 3D charge density in a unit cell of the untwisted bilayer BP. (b) 3D charge density in a unit cell of 70.53° twisted bilayer BP. (c) 2D charge density map of untwisted bilayer BP. The cutting plane (020) is the middle-distance plane between the two 2D BP atomic planes. (d) 2D charge density map in the same cutting plane for the 70.53° twisted bilayer BP. (e) 2D charge density map for another cutting plane, (100) of twisted bilayer BP, which is normal to the 2D atomic planes.